\documentclass[epj]{svjour}

\usepackage{graphics}
\begin{document}
\title{Same Sign WW Scattering Process as a Probe of Higgs Boson in pp
  Collision at $\sqrt{s}$~=~10~TeV}
\author{
  Bo Zhu\inst{1}
  \and Pietro Govoni\inst{2,3,4}
  \and Yajun  Mao\inst{1}
  \and Chiara Mariotti\inst{5}
  \and Weimin Wu\inst{6}
\thanks{Supported by the National Natural Science Foundation of
China (10099630), Ministry of Science and Technology of
China(2007CB816101) and China Scholarship Council.}
}                     % Do not remove
\institute{ School of Physics, and State Key Laboratory of Nuclear
Physics \& Technology, Peking University, China \and Milano-Bicocca
University, Italy \and INFN Milano-Bicocca, Italy \and European
Organization for Nuclear Research \and INFN Torino, Italy \and Fermi
National Accelerator Laboratory, Batavia, IL, USA }

\date{Received: date / Revised version: date}
\abstract{ WW scattering is an important process to study
electroweak symmetry breaking in the Standard Model at the LHC, in
which the Higgs mechanism or other new physics processes must
intervene to preserve the unitarity of the process below 1 TeV. This
channel is expected to be one of the most sensitive to determine
whether the Higgs boson exists. In this paper, the final state with
two same sign Ws is studied, with a simulated sample corresponding
to the integrated luminosity of 60 fb$^{-1}$ in pp collision at
$\sqrt{s}=$10 TeV. Two observables, the invariant mass of $\mu\mu$
from W decays and the azimuthal angle difference between the two
$\mu$s, are utilized to distinguish the Higgs boson existence
scenario from the Higgs boson absence scenario. A good signal
significance for the two cases can be achieved. If we define the
separation power of the analysis as the distance, in the
log-likelihood plane, of pseudo-experiments outcomes in the two
cases, with the total statistics expected from the ATLAS and CMS
experiments at the nominal centre-of-mass energy of 14 TeV, the
separation power will be at the level of 4~$\sigma$.
% $\alpha=\frac{N_{NoH}-N_{m_{H}(200)}}{\sqrt{N_{m_{H}(200)}+N_{Bkg}}}$.
%
 % end of PACS codes
} %end of abstract
\maketitle
\section{Introduction}
\label{sec:intro}

It is predicted by the Standard Model(SM) that perturbative
unitarity is violated in vector boson scattering process at high
energy if the Higgs particle is absent\cite{EWSB}. This implies that
the existence of a Higgs boson or new physics must intervene below
1 TeV. If the Higgs boson does exist, a resonance could be observed
in the VV (WW or ZZ) invariant mass spectrum. On the other hand, new
physics may appear in the form of vector boson pair resonances, as
predicted by Little Higgs, Dynamical symmetry breaking, or Higgsless
models\cite{EWSB}. Therefore, a measurement of WW scattering
processes is a model independent approach to probe the existence or
absence of a Higgs boson.

\begin{figure}
\begin{center}
\resizebox{0.3\textwidth}{!}{%
  \includegraphics{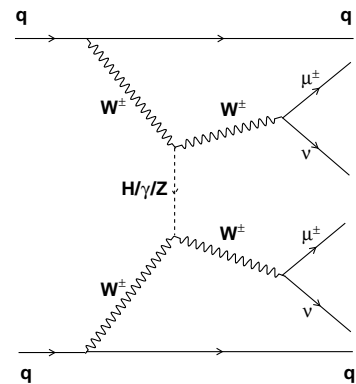}
}
%\vspace{4.2cm}       % Give the correct figure height in cm
\caption{\quad Same Sign WW Scattering Diagram.}
\label{fig:diagram}       % Give a unique label
\end{center}
\end{figure}

The same sign WW scattering with W decaying to $\mu\nu$ is expected
to be a very clean process to study the difference between the
standard model and new physics scenarios \cite{accomando}. It has
the best separation power between the two scenarios with respect to
the other final states (WW, ZZ, WZ) as shown in \cite{accomando}. It
will help clarify the electroweak breaking mechanism in case a Higgs
boson like resonance will not be observed or to finally test the
unitarity of the theory. A characteristic signature of the same sign
WW scattering is the presence of two forward jets (tag jets) with
high energy (see Fig.1) which can thus be efficiently extracted from
most backgrounds. The other signature, namely the presence of a same
sign isolated muons pair, can help in suppressing other backgrounds.
In this work, we take into account all the possible backgrounds,
including that due to the mis-identification of leptons (which is
usually neglected in other same sign WW scattering studies). We will
show that we  can get an almost background free result with the help
of isolation techniques. The final state with 2 electrons or 1
electron and 1 muons have been studied, but the background
subtraction result is much less effective, due to the high rate of
mis-identified electrons.

%It should be mentioned that there is no WW resonances.
Two same sign WW are produced only via t-channel process, thus no
resonances are expected in the $m_{WW}$ spectrum. The invariant mass
of the WW is shown in Fig.2 at parton level for two different values of the Higgs
boson mass and for the case of no-Higgs. Because of
the Parton Distribution Functions, the expected rise at large
$m_{WW}$ values is dramatically suppressed, but still a substantial
difference between the two scenarios (Higgs and ``no-Higgs'') can
clearly be observed.

%no VV resonances exist and the two final state neutrinos bring large
%missing energy, which indicates that we can't get invariant mass of
%W bosons.

\begin{figure}
\begin{center}
\resizebox{0.45\textwidth}{!}{%
  \includegraphics{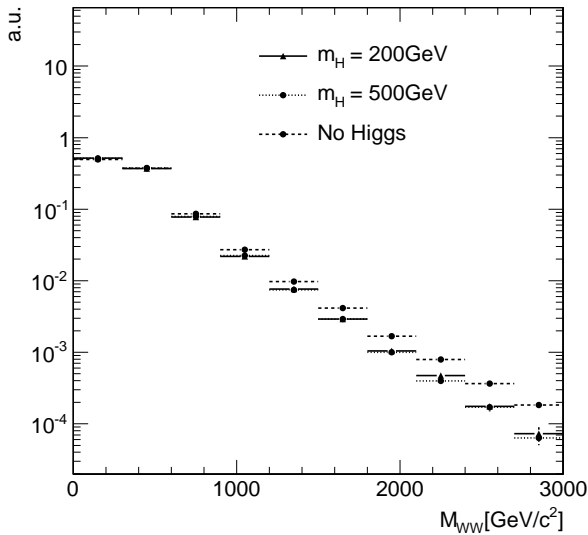}
}
\vspace{0.2cm}       % Give the correct figure height in cm
\caption{\quad $m_{WW}$ distribution for $m_H$~=~200~GeV$/c^{2}$,
$m_H$~=~500~GeV$/c^{2}$ and no-Higgs Scenarios. The distribution is
normalized to 1.} \label{fig:mww}
\end{center}
\end{figure}

\section{Monte Carlo Samples}
\label{sec:samples}

The PHANTOM events generator \cite{phantom} is used to generate
$qq\rightarrow qq\mu^{\pm}\nu\mu^{\pm}\nu$ processes at
$\mathcal{O}(\alpha_{EW}^6)$ , since it performs the full
calculations at $\mathcal
{O}(\alpha_{EW}^6+\alpha_{EW}^4\alpha_{S}^2)$ order. This is
necessary, since the study aims at comparing the WW scattering
spectra under two different Higgs boson hypotheses: thus it is of
crucial importance to correctly calculate the cross sections, by
considering the interferences between the various tree-level
diagrams present in the WW scattering process calculation.

%The parton shower and hadronization are simulated with
%PYTHIA\cite{pythia}.
Different Higgs boson hypotheses samples are generated for the
signal: $m_H$~= 200 GeV$/c^{2}$, $m_H$ = 500 GeV$/c^{2}$ and
no-Higgs scenarios. Out of all the possible diagrams calculated by
PHANTOM, the WW scattering process is isolated by means of the
following cuts at parton level: the invariant mass constraint
$|m_{\mu\nu}-m_W|$~$<$ 10 GeV$/c^{2}$, the pseudo-rapidity
difference of the final state quarks $\Delta \eta_{qq}$~$>$~2.0, the
invariant mass of the quarks $m_{qq}$ $>$ 300 GeV$/c^{2}$, the
minimal angle between the final state muon and quark $\Delta R(\mu
q)^{min}$ $ >$ 1.2. After these selections surviving events are
considered as signal events, the remaining events are studied as
irreducible background.

Besides the irreducible background, some other processes at
$\mathcal {O}(\alpha_{EW}^4\alpha_{s}^2)$~\cite{phantom} with the
same final states particles are also produced by PHANTOM. These
processes are denoted as ``QCD background'' in the following.

The $t\bar{t}\rightarrow W^{+}bW^{-}b$ production is another very
important background, in which one hard muon comes from W, the other
same sign muon is from a b-hadron leptonic decay. Single top quark
in association with W process is also considered because of the same
reason.

The production of single W along with jets, in which the W decays
into $\mu\nu$ is another dangerous background, because charged
long lived hadrons $(k^{\pm}$,$\pi^{\pm}$,$p^{\pm}$) may be wrongly
identified as muons, and the large cross section compensates for the
low probability of the mis-identification. We assume the probability
of mis-identification to be $5\times10^{-4}$~\cite{cmsptdr}. In
addition to the dominant backgrounds discussed above, single top,
$t\bar{t}$W and di-boson backgrounds (WW, WZ and ZZ) are studied as
well.

%The cross sections of signal and background processes are shown in
%Table ~\ref{tab:sigbkg}.

%\textbf{I WOULD SAY FOR EACH BACKGROUND, WHICH GENERATOR IS USED.}
QCD and irreducible background samples are produced with PHANTOM,
$t\bar{t}$, W+jet and $t\bar{t}$W backgrounds are generated with
Madgraph\cite{madgraph} and the other backgrounds are simulated with
PYTHIA at a collision energy of $\sqrt{s}=$10 TeV. The cross
sections of the samples which are produced by PHANTOM are calculated
at the Leading Order (LO), the cross sections of the other samples
are calculated at the Next-to-Leading Order (NLO). The cross section
will be roughly doubled if the collision energy is raised from 10
TeV to 14 TeV. In all cases including signal and background samples,
the parton showering and hadronization are performed with PYTHIA,
and the jet reconstruction algorithm is also provided by PYTHIA. To
include the detector effect, the muons and jets momenta are smeared
by a gaussian distribution with the resolution based on the
following $p_T$ resolution parameterization\cite{resolution},

%The parton shower and hadronization are simulated with
%PYTHIA\cite{pythia}.

for muons:
\begin{eqnarray}
\label{muonsmear} \frac{\sigma(p_{T})}{p_{T}}=e^{-4+0.0014\times
p_T};
\end{eqnarray}

for jets:
\begin{eqnarray}
\label{jetsmear}
\frac{\sigma(p_{T})}{p_{T}}=\sqrt{\frac{0.813^2}{p_{T}}+\frac{3.9^2}{p^2_{T}}+0.017^2}.
\end{eqnarray}

\section{Event Selection}
\label{sec:selection}

The aim of the selection strategy is to achieve a
reasonable level of signal over background ratio. We concentrate on
a cut-based selection strategy. The selection chain includes two
main parts: muon selection and jet selection.

%\subsection{muon Cut}

A pair of same sign isolated hard muons is one of the most
significant characteristics of the signal process. Most standard
model background events, such as W+jet, $t\bar{t}$, single top and
di-boson, comprise only one muon or two opposite charged muons in
the final state. If there are two same sign muons in these events,
one muon should come from b-hadron decay or muon mis-identification
from other backgrounds. %Besides the muons sign, their $p_T$ value and isolation
%are used to constrain the backgrounds.
Most of the non-top background events contain at least one fake muon
mostly in the low $p_T$ region. A $p_T$ threshold of 15~GeV$/c$ is
required to suppress these kinds of background, especially the W+jet
events.

The muon isolation criteria are applied to all the tracks of charged
particles, which can be well reconstructed with an efficiency of
almost $100\%$ when $p_T$~$>$~0.5~GeV$/c$~\cite{resolution}. The
isolation parameter is defined as the sum of the $p_T$ of charged
particles in an isolation cone of 0.3~rad centered around the muon
at the primary vertex, in the ($\eta$,$\phi$) plane. The footprint
of the muon itself is removed by an inner veto cone of 0.01~rad:
\begin{eqnarray}
\label{isolation} \beta={\Sigma p_{T}(0.01<\Delta R<0.3)}.
\end{eqnarray}
As the top background is the most important one, the following
isolation cuts are tuned to reduce this contribution:
$\beta<$1~GeV$/c$ and $\beta/p_T(\mu)$~$<$~0.05.

%\subsection{Forward Jet Cut}

The vector boson scattering signature is exploited as well to
further reduce the backgrounds contribution. The tag jets are
identified as the ones with highest $p_T$ in the event.
There will be very high fake rate for low $p_T$ jets,
so the $p_T$ threshold of the tag jets is 30~GeV$/c$. %Many different
%strategies are possible for implementing a tag-jet selection.
A number of different strategies to implement tag jets selection
were compared, and the best rejection factor for a given efficiency
is obtained by requiring the tag jets with the opposite sign of
pseudo-rapidity ($\eta$), to satisfy the $\eta$ difference $\Delta
\eta_{jj}$~$>$~4 and tag jets invariant mass
$m_{jj}$~$>$~600~GeV$/c^{2}$.

%As it is done for the selections at parton level, muons are required
%to be central and far away from tag jets to reduce the irreducible
%background.

The event number after the cut-based selection for signal and
background are shown in Table 1. The results are normalized to an
integrated luminosity of 60~fb$^{-1}$. For $t\bar{t}$, W+jet, single
top and di-boson backgrounds, Monte Carlo samples corresponding to
60~fb$^{-1}$ are too large to be simulated, due to the very large
cross section. Only few events survive after the selection chain
with high statistics error. The expected number of events therefore
will be estimated with the efficiency factorization as discussed
below.

\begin{table}
\label{tab:survived}\begin{center}
\begin{tabular}{ccc}
\hline\noalign{\smallskip}
  $m_H$~=~200~GeV$/c^{2}$            & no-Higgs      & Backgrounds \\
 \noalign{\smallskip}\hline\noalign{\smallskip}
  12.2                               & 13.7          & 5.9        \\
\noalign{\smallskip}\hline
\end{tabular}%
\vspace*{0.3cm} \caption{Number of surviving events for signal and
background after muon and jet selection with an integrated
luminosity of 60~fb$^{-1}$}
%\textbf{WRITE IN THE CAPTION THE LUMINOSITY THEY REFER TO.
\end{center}
\end{table}

%\begin{table}
%\label{tab:survived}\begin{center}
%\begin{tabular}{cccc}
%\hline\noalign{\smallskip}
% Signal:     & $m_H$~=~200~GeV$/c^{2}$ &           & no-Higgs  \\
% \noalign{\smallskip}\hline\noalign{\smallskip}
%             & 12.2                    &           & 13.7       \\
% \noalign{\smallskip}\hline\noalign{\smallskip}
%Background:  & QCD+Irr                 & $t\bar{t}$& $t\bar{t}$W \\
% \noalign{\smallskip}\hline\noalign{\smallskip}
%             & 1.7                     & 4         & 0.2 \\
%\end{tabular}%
%\vspace*{0.3cm} \caption{Number of surviving events for signal and
%background after muon and jet selection with an integrated
%luminosity of 60~fb$^{-1}$}
%%\textbf{WRITE IN THE CAPTION THE LUMINOSITY THEY REFER TO.
%\end{center}
%\end{table}

\section{Higgs versus no-Higgs scenario}
\label{sec:distingish}

To distinguish the scenario where the Higgs boson is existing from
the one where the Higgs boson is absent, two possible additional
selections have been investigated. We choose the following relative
separation definition to optimize the selections:
\begin{eqnarray}
\label{mysignificance}
\alpha=\frac{N_{NoH}-N_{m_{H}(200)}}{\sqrt{N_{m_{H}(200)}+N_{Bkg}}}.
\end{eqnarray}
where $N_{m_{H}(200)}$, $N_{NoH}$ and $N_{Bkg}$ are the number of
events for the two cases and for the backgrounds respectively. For
this study the value of the Higgs boson mass is not relevant, as
explained in detailed in ref.\cite{accomando}

The region of high values of invariant mass of W bosons $(m_{WW})$
should be sensitive to the presence of a Higgs particle (Fig.2).
Unfortunately, because of the presence of neutrinos, it is
impossible to reconstruct the invariant mass of the W bosons.
Therefore, the invariant mass of the two muons system is used to
replace $m_{WW}$ and the events count for Equation 4 is performed
after a cut on the $m_{\mu\mu}$ value.
%\subsection{High Invariant Mass Result}

Fig.3 (a) shows the $m_{\mu\mu}$ distribution for the two scenarios
($m_{Higgs}$~=~200~GeV$/c^2$ and no-Higgs). Fig.3 (b) shows the
number of surviving signal events as a function of the $m_{\mu\mu}$
cut. Fig.3 (c) shows the distribution of the relative separation (as
defined in Equation 4) $vs.$ the cut on $m_{\mu\mu}$. To obtain a
better separation between the two cases, we require the muon to be
in the central region: $|\eta_{\mu}|$~$<$~2. By asking
$m_{\mu\mu}$~$>$~200~GeV$/c^2$, we can achieve good signal
significance and background control. However, the request is too
tight, since it eliminates about 80$\%$ of signal events (Fig.3
(b)).

\begin{figure*}
\vspace*{0.1cm}
\resizebox{1.0\textwidth}{!}{%
  \includegraphics{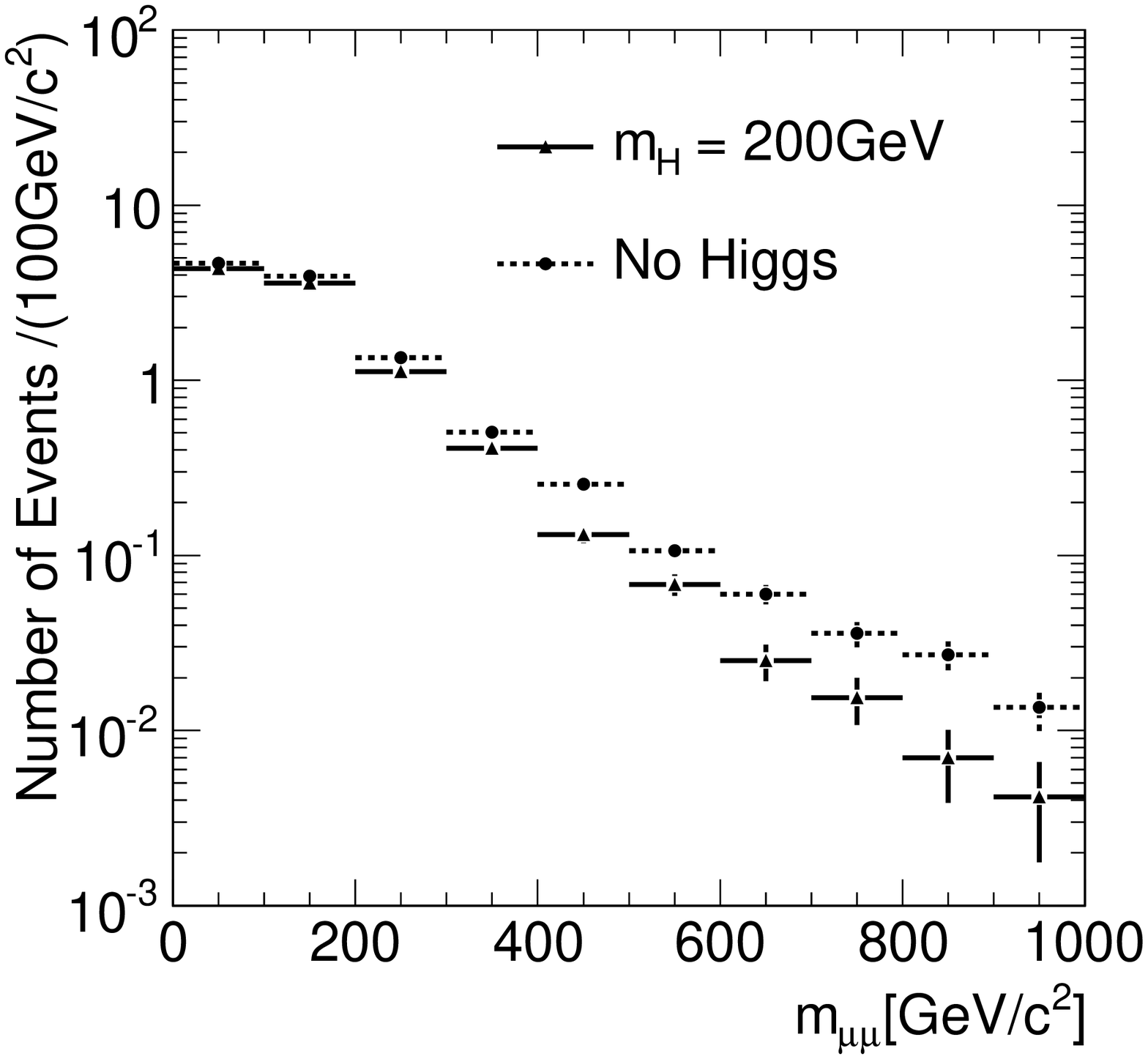}
  \includegraphics{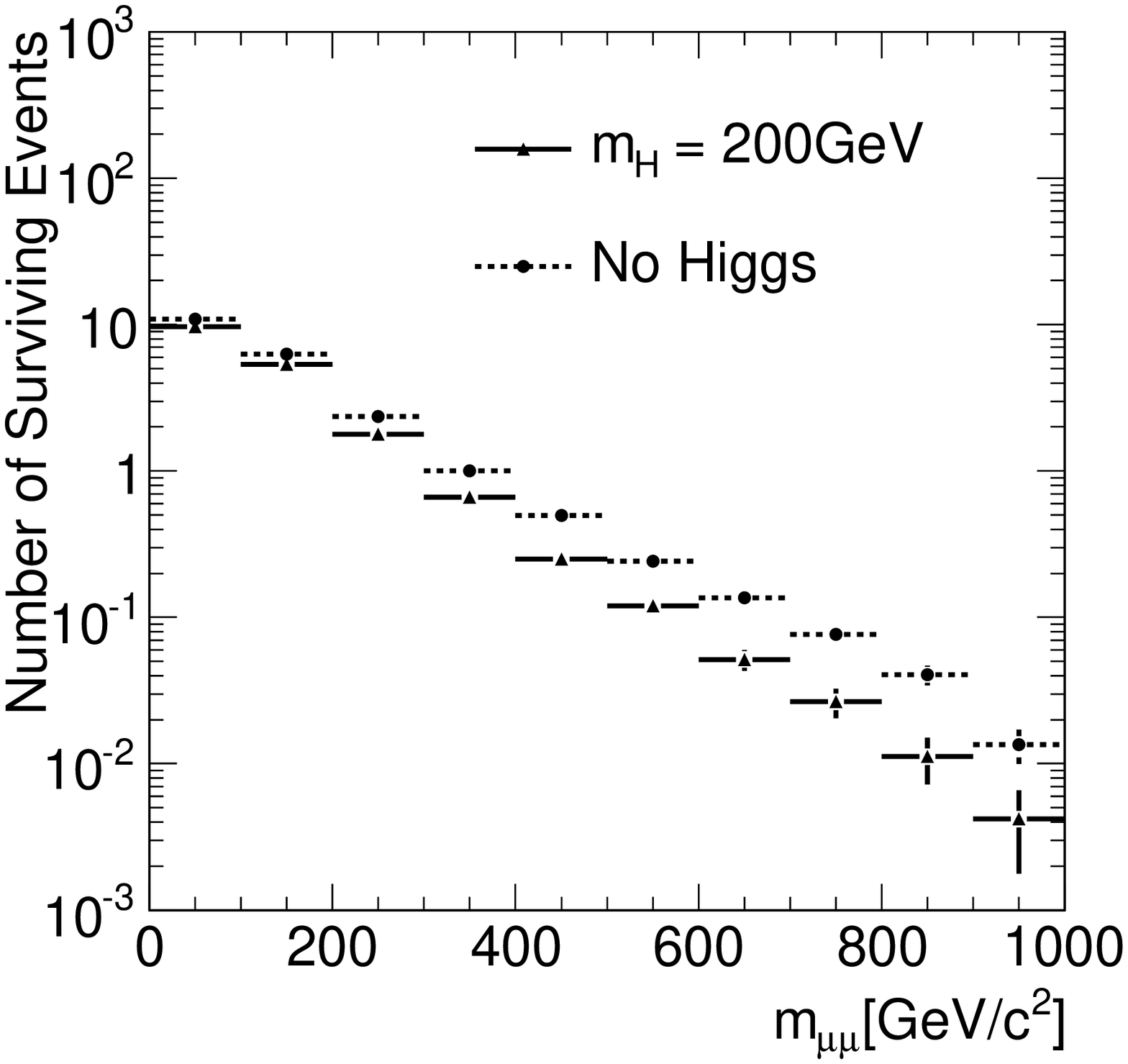}
  \includegraphics{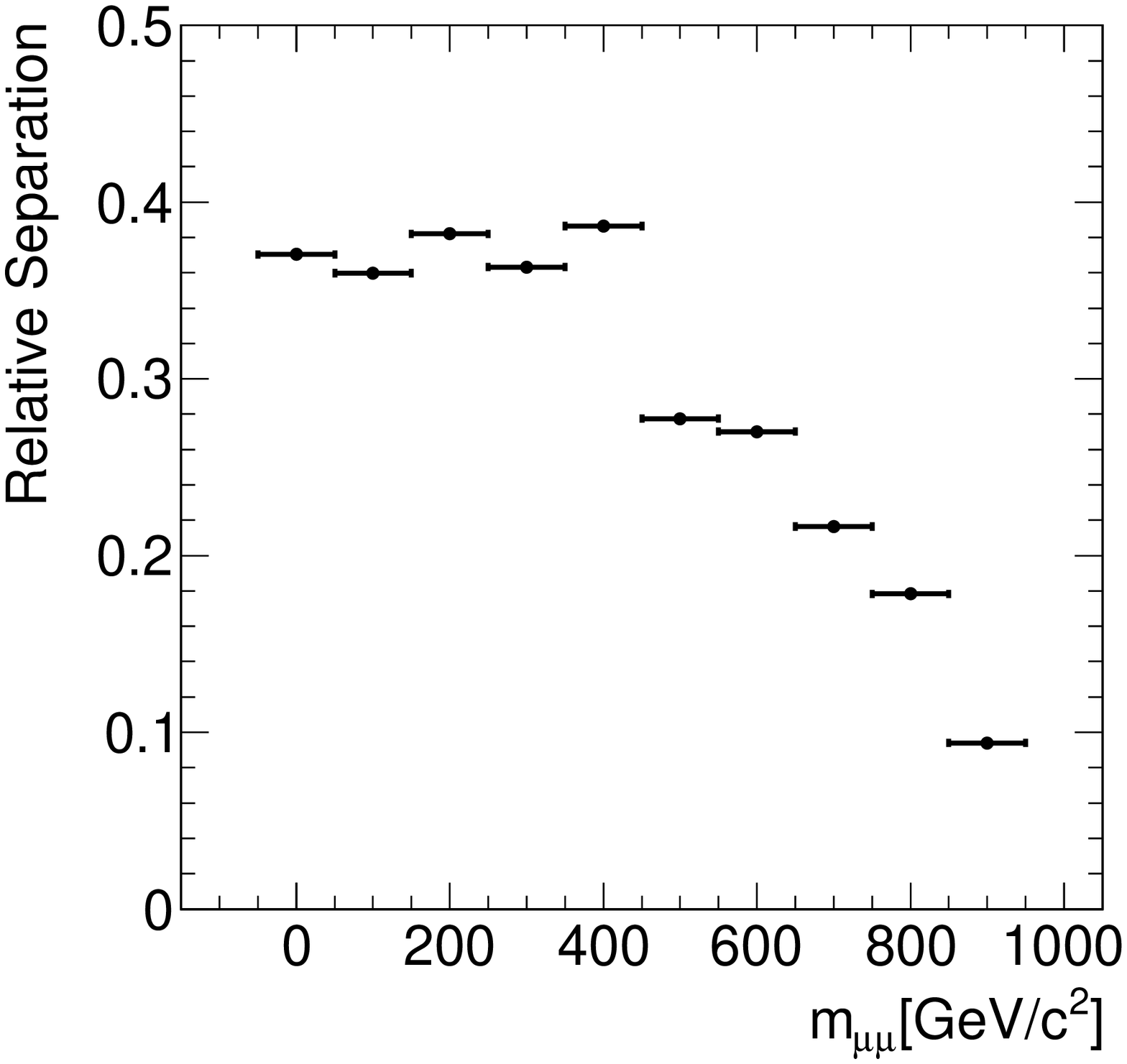}\\
}
\begin{center}{
 a
 ~~~~~~~~~~~~~~~~~~~~~~~~~~~~~~~~~~~~~~~~~~~~~~~~~~~~~ b
 ~~~~~~~~~~~~~~~~~~~~~~~~~~~~~~~~~~~~~~~~~~~~~~~~~~~~~ c\\
 }
 \end{center}
 \caption{Invariant mass distribution of the two muons ($m_{\mu\mu}$) (a) ,
      the number of surviving events as a function of the cut on $m_{\mu\mu}$ (b),
      relative separation $\alpha$ $vs.$ the $m_{\mu\mu}$ cut value (c). Results are
      normalized to 60~fb$^{-1}$.} \label{fig:mass}
\end{figure*}

Alternatively, a selection on the azimuthal angle between muons is
investigated, as the vector bosons tend to be back to back in a
scattering topology. Fig.4 (a) shows the $\Delta\phi$ distribution
between the two muons for the two cases. Fig.4 (b) shows the number
of surviving events as a function of a minimum $\Delta\phi_{\mu\mu}$
cut. Fig.4 (c) is the distribution of the relative separation as
defined in Equation 4 $vs.$ different $\Delta\phi_{\mu\mu}$ cuts.
With the cut $\Delta\phi_{\mu\mu}$~$>$~2, the highest separation is
obtained with a loss of about 50$\%$ of signal events. Only QCD and
irreducible backgrounds are considered in Fig.3 (c) and Fig.4 (c).

\begin{figure*}
\vspace*{0.1cm}
\resizebox{1.0\textwidth}{!}{%
  \includegraphics{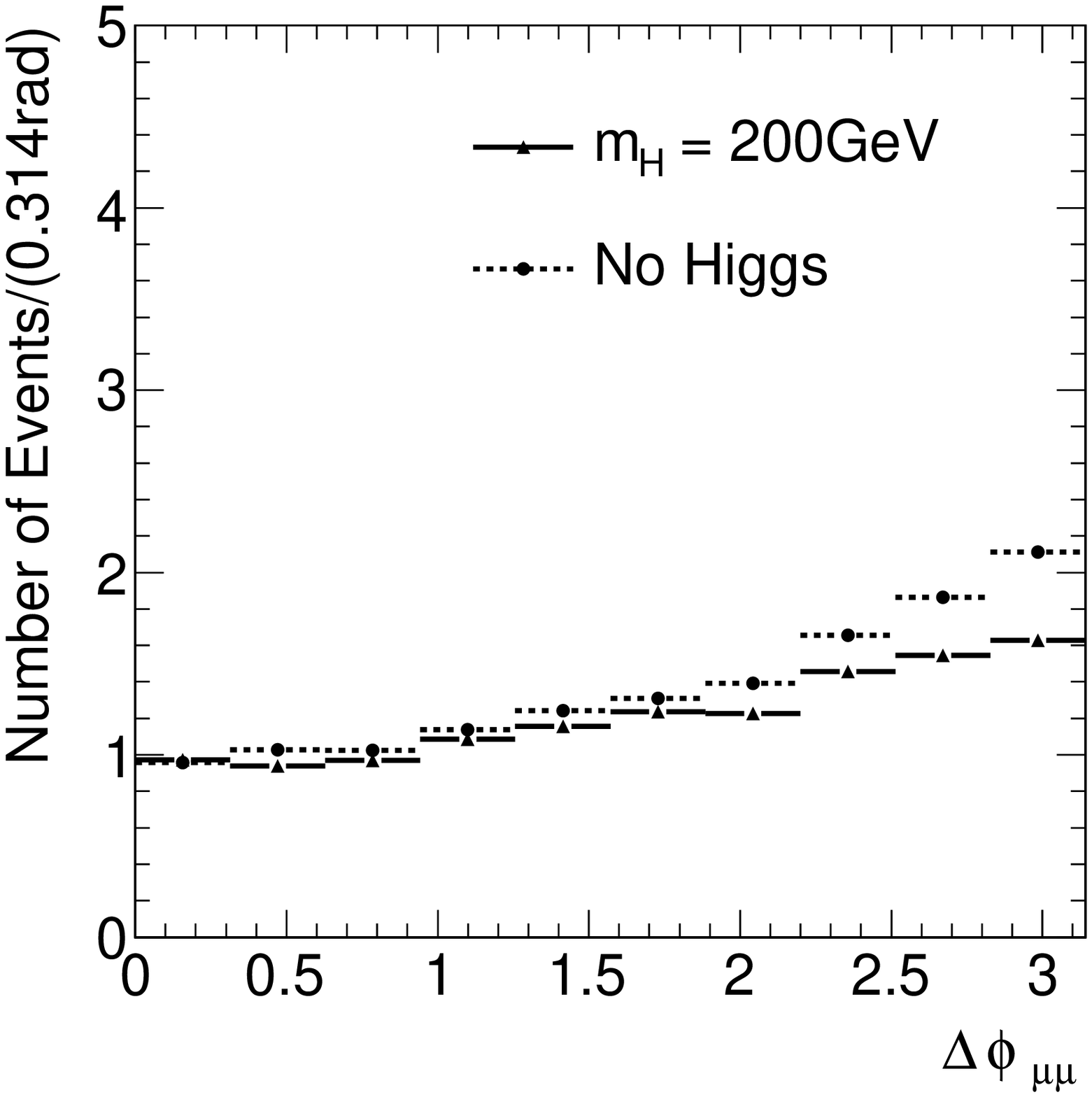}
  \includegraphics{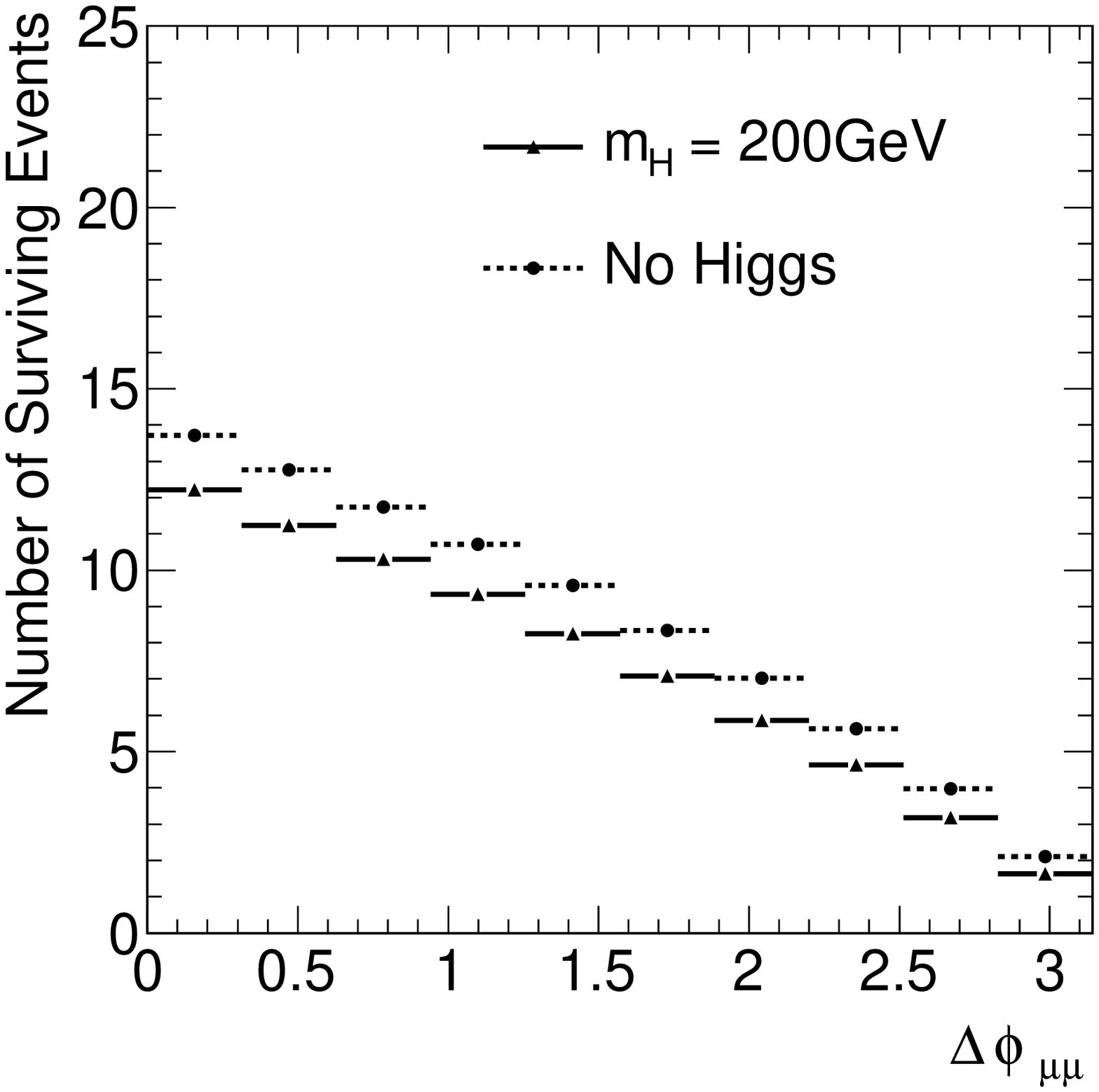}
  \includegraphics{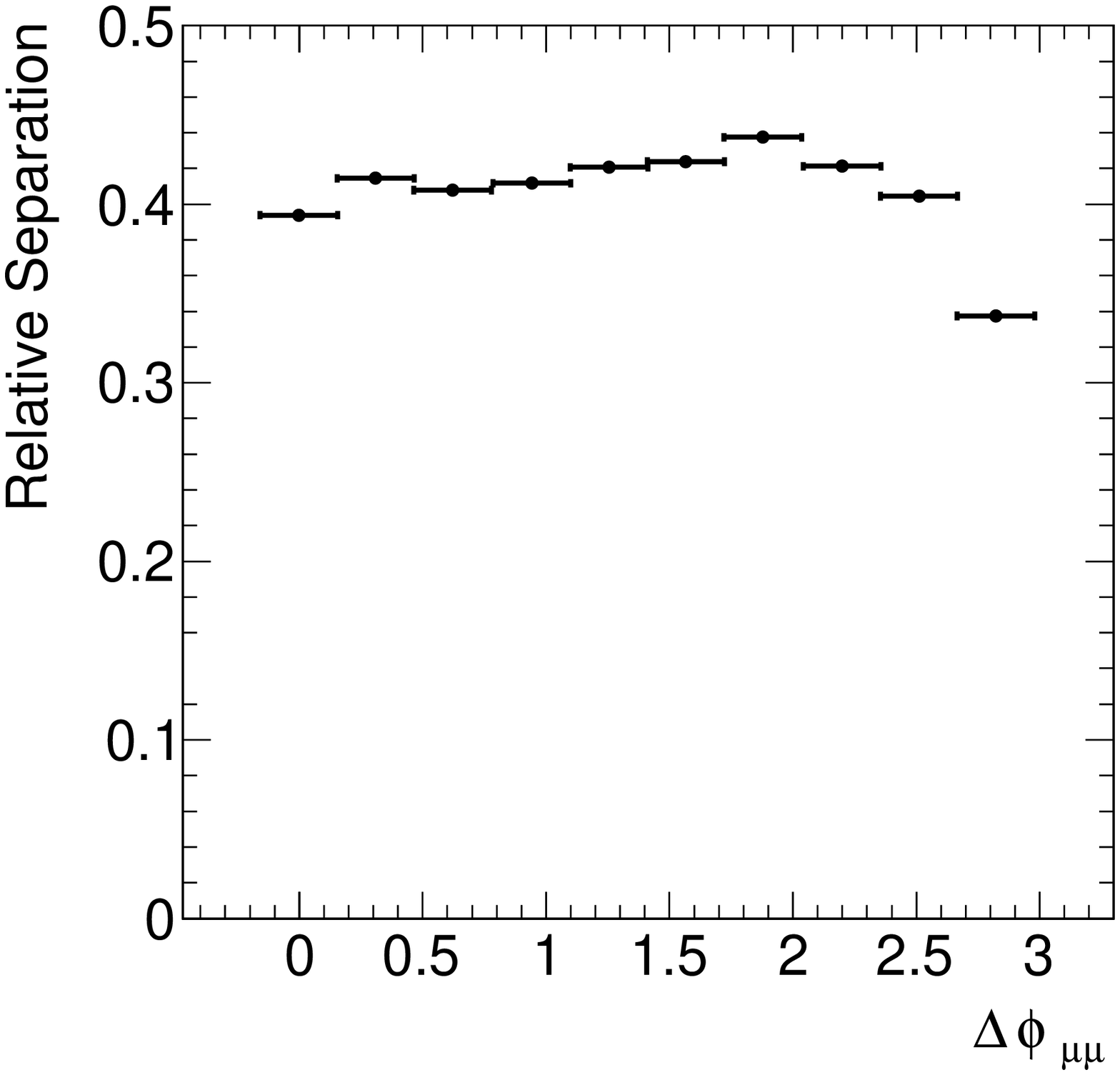}\\
}
\begin{center}{
 a
 ~~~~~~~~~~~~~~~~~~~~~~~~~~~~~~~~~~~~~~~~~~~~~~~~~~~~~ b
 ~~~~~~~~~~~~~~~~~~~~~~~~~~~~~~~~~~~~~~~~~~~~~~~~~~~~~ c\\
 }
  \end{center}
 \caption{$\Delta\phi_{\mu\mu}$ distribution (a),
      the number of surviving events as a function of the  $\Delta\phi_{\mu\mu}$ cut (b),
     relative separation $\alpha$ $vs.$  the $\Delta\phi_{\mu\mu}$ cut value
      (c). Results are normalized to 60~fb$^{-1}$.  } \label{fig:dphi}
\end{figure*}

\section{Background estimation}
\label{sec:bkg}

The main uncertainty comes from the simulated background samples
statistical error. Because of the limited statistics available, no
event remains for W+jet, top and di-boson samples. However, we
cannot ignore those backgrounds because of their very large cross
sections.

Assuming there is no correlation among the single selections, we
estimate the number of surviving events by multiplying the single
efficiencies:
\begin{eqnarray}
\label{mysignificance} N=\sigma\times L(60 fb^{-1})\times
\xi_{cut1}\times \xi_{cut2}...\times \xi_{cuti},
\end{eqnarray}
where the $\xi_{cuti}$ is the efficiency for the i-th selection
alone on each sample. There is a very low level of correlation
between the two main selections, namely the jet selections and muon
selections. The expected number of background events for each sample
using two different discriminators are summarized in Table 2 with an
integrated luminosity of 60 fb$^{-1}$ .

\begin{table}
\label{tab:estimate} \begin{center}
\begin{tabular}{c|ccc}
\hline\noalign{\smallskip}
 Discriminator               &top        & W+jet       & di-boson    \\
 \noalign{\smallskip}\hline\noalign{\smallskip}
  $m_{\mu\mu}$               & 0.65      &  0.05       & 0.02             \\
  $\Delta\phi_{\mu\mu}$      & 2.6       &  0.2        & 0.1                 \\
\noalign{\smallskip}\hline
\end{tabular}%
\vspace*{0.1cm} \caption{Estimated number of events of backgrounds}
\end{center}
\end{table}
%\vspace{0.3cm}

The signal significance is determined using the likelihood ratio
method, with poissonian probability density distributions, for both
the $m_{\mu\mu}$ and $\Delta\phi_{\mu\mu}$ selections with the
background estimates in Table 2. Results are listed in Table 3. The
number of signal and background events are shown after the
selection. We make the hypothesis that the correlation between the
cuts will give 100$\%$ uncertainty for W+jet, top and di-boson
backgrounds. For the other samples, only the statistical error is
considered.

\begin{table*}
\label{tab:everyth} \begin{center}
\begin{tabular}{c|ccccccc}
\hline\noalign{\smallskip}
 Discriminator  & No H       & $m_{H}(200)$ & ${NoH}/{m_{H}(200)}$ & Background  & Relative Separation & $S_{(m_{H}(200))}$& $S_{No H}$\\
\noalign{\smallskip}\hline\noalign{\smallskip}
 $m_{\mu\mu}$          & 2.4$\pm$0.1 & 1.8$\pm$0.1 & 1.3$\pm$0.1    & 1.0$\pm$0.8 & 0.35         & 1.5 & 1.9 \\
 $\Delta\phi_{\mu\mu}$ & 6.5$\pm$0.1 & 5.4$\pm$0.1 & 1.2$\pm$0.1    & 3.5$\pm$2.9 & 0.36         & 2.4 & 2.8 \\
\noalign{\smallskip}\hline
\end{tabular}%
\vspace*{0.3cm} \caption{Signal significances, ratio and separation
power between Higgs case and no-Higgs case with an integrated
luminosity of 60 fb$^{-1}$, at 10~TeV centre-of-mass energy.}
\end{center}
\end{table*}

\section{Summary and Discussion}

\begin{figure}
\begin{center}
\resizebox{0.45\textwidth}{!}{%
  \includegraphics{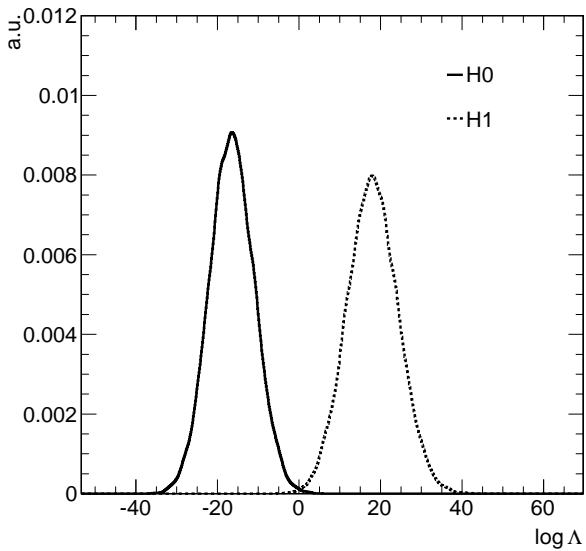}
}
%\vspace{4.2cm}       % Give the correct figure height in cm
\caption{\quad Normalized Likelihood Ratio with $m_{\mu\mu}$ cut, H0
hypothesis is $m_H$~=~200GeV, H1 is no higgs. Result is
corresponding to an inverse luminosity of 6~ab$^{-1}$ at
$\sqrt{s}$~=~14~TeV } \label{fig:LLR}
\end{center}
\end{figure}

Assuming a poissonian pdf of the measurements in the Higgs boson existing
scenario and Higgs boson absence scenario, a likelihood-ratio is built  to
distinguish the two hypotheses, giving the
number of measured events.
To assess the separation power of the analysis, a set of toy-montecarlo experiments have been generated
for each of the cases, and the distributions of the corresponding
likelihood-ratios have been compared. To evaluate the separation between
the two curves, the distance between their maxima, normalized to
their sigma, is calculated (the sigma is taken as the half-width of
the narrowest interval containing 68\% of the distribution):
\begin{equation}
\delta~=~\frac{|\max(LLR)_{H0} -
\max(LLR)_{H1}|}{\sigma_{H0}~\oplus~\sigma_{H1}}
\end{equation}
where $H0$ and $H1$ represent the Higgs boson and no-Higgs boson hypotheses
respectively.  Fig.\ref{fig:LLR} shows the distributions when
we scale the results by the total expected statistics
collected by ATLAS and CMS  (corresponding to an inverse
luminosity of 6~ab$^{-1}$ at 14~TeV of LHC centre-of-mass energy).
A 4$\sigma$ separation for the two hypotheses can be achieved.

We present an exploratory study of the same sign W scattering
process with W decay into $\mu\nu$ as probe of Higgs boson existence
in pp collisions at $\sqrt{s}$~=~10~TeV. All the standard model
backgrounds are considered, with detector effects parameterized,
including muon mis-identification effect. It is a clean channel
compared with the other VV scattering
processes~\cite{accomando}\cite{vvscat} because of the two main
signatures, which are the same sign isolated muons pair and
energetic forward jets. $m_{\mu\mu}$ and $\Delta\phi_{\mu\mu}$ are
both good discriminants to distinguish a Higgs scenario from the
no-Higgs one.

Although the cross section is not as large as searching for Higgs
Boson via di-boson resonances directly, it is a model independent
channel to determine if the Higgs boson exists, whatever the value
of its mass, and to verify the unitarity of the theory.

With the total statistics expected from the ATLAS and CMS
experiments at 14 TeV, the separation power between Higgs boson
and no-Higgs boson scenarios will be at the level of
4~$\sigma$.

%
% No-BibTeX users please se


\begin{thebibliography}{}
%Author, \textit{Book title} (Publisher, place year) page numbers
% etc

\bibitem{EWSB}  M.J.G. Veltman, CERN-97-05; M.S. Chanowitz, [hep-ph/9812215]; S.
Dawson, [hep-ph/9901280]; Chris Quigg, Acta Phys.Polon. B30 (1999)
2145. [hep-ph/9905369]; S. Dawson Int.J.Mod.Phys. A21 (2006) 1629.
[hep-ph/0510385]; R. Rattazzi PoS HEP2005 (2006)
399.[hep-ph/0607058]

\bibitem{accomando} E. Accomando et al., JHEP 0603 (2006) 093;
\bibitem{bagger} J. Bagger et al Phys.Rev.D52:3878-3889,1995.

\bibitem{phantom} A.Ballestrero, A.Belhouari, G.Bevilacqua etc.
Computer Physics Communications 180 (2009) 401¨C417

%\bibitem{accomando} E.Accomando et al. "Boson fusion and Higgs
%production at the LHC in six fermion final state with one charged
%lepton pair" arXiv:hep-ph/0603167v4

\bibitem{pythia} T.Sjostrand, S.Mrenna and P.Skands, JHEP 0605 (2006) 026

\bibitem{cmsptdr} The CMS Collaboration, CERN-LHCC-2006-001

%\bibitem{wwnlo} T.Melia, K.Melnikov, R.Rontsch, G.Zanderighi,[arXiv:1007.5313v1]


\bibitem{madgraph} F. Maltoni and T. Stelzer, JHEP 0302 (2003) 027
[arXiv:hep-ph/0208156]; T. Stelzer and W. F. Long, Comput. Phys.
Commun. 81 (1994) 357 [arXiv:hep-ph/9401258].

\bibitem{resolution} The ATLAS Collaboration, CERN-LHCC-1999-14 vol 1

\bibitem{vvscat} A.Sznajder for CMS Collaboration,
[arXiv:0810.3604v2]


\end{thebibliography}
\end{document}